\documentclass[showpacs,amsmath,amssymb,twocolumn,floatfix,prl]{revtex4}
\usepackage[dvips]{graphicx}
\usepackage{amssymb}
\input{epsf}

\begin{document}

\title{
Dynamical localization, measurements  and quantum computing
}


\author{M. Terraneo and D. L. Shepelyansky}
\affiliation{Laboratoire de Physique Th\'eorique, UMR 5152 du CNRS, 
Univ. Paul Sabatier, 31062 Toulouse Cedex 4, France}

\date{September 26, 2003}

\begin{abstract}
We study numerically the effects of measurements on  dynamical localization
in the kicked rotator model simulated on a quantum computer. Contrary to the
previous studies, which showed that measurements induce a diffusive probability
spreading, our results demonstrate that localization can be preserved 
for repeated single-qubit measurements. We detect a transition from 
a localized to  a delocalized phase, depending on the system parameters and on 
the choice of the measured qubit. 
\end{abstract}

\maketitle

In 1979 the dynamical localization of quantum chaos was 
discovered in numerical simulations of the kicked rotator model \cite{casati}. 
It was found that the unbounded classical diffusion 
typical of chaotic dynamics is suppressed  by quantum interference effects 
\cite{casati,chirikov}.
This interesting phenomenon found its explanation on the basis of  an
analogy with the Anderson localization in disordered lattices 
\cite{fishman} (see also \cite{reviews}).  
Manifestations of dynamical localization appear in various physical systems. 
Its first experimental observation was obtained
with hydrogen and Rydberg atoms in  a microwave field \cite{koch}.
Recently, a significant technological progress in manipulating
cold atoms by laser 
fields allowed to build up experimentally the kicked rotator model and to study
dynamical localization in real systems in great detail 
\cite{raizen,newzealand,oxford}.

Since localization appears due to quantum interference it is natural to expect
that it is rather {\it fragile} and sensitive to noise and interactions with
the environment.  Indeed, in theoretical and experimental studies 
it was shown  that even a small amount of noise destroys coherence and 
localization \cite{ott,newzealand,oxford}. Measurements  represent
another type of coupling to the environment \cite{zurek}, 
and it is of fundamental  importance to understand their 
effects on dynamical localization.
Theoretical and numerical studies show that measurements destroy 
localization and induce a diffusive energy growth like in the case of  a noisy
environment 
\cite{graham,kaulakys,facchi}. 
For weak continuous measurements, discussed in \cite{graham},  the rate of this
growth can be much smaller than the diffusion rate induced by classical chaos.
However in the limit of strong coupling to the measurement device the quantum
diffusion rate becomes close to its classical value. A similar situation
takes place in the case of projective measurements, considered in 
\cite{kaulakys, facchi}. 

The interest in  measurement procedures enormously grew up in the last years
due to progress in quantum information processing \cite{chuang}. Indeed, the
extraction of information from a quantum computation is always reduced to a 
final measurement of the quantum register. Various experimental implementations
were discussed for the realization of the readout procedure in quantum 
optics systems \cite{cirac,wiseman} and
solid state devices \cite{gurvitz,korotkov,averin,devoret}. 
Moreover it has been shown
 that a quantum computation can be performed completely by a sequence of
measurements applied to an initially entangled state \cite{briegel}. 
At the same time, measurements represent an important part of various 
quantum algorithms, including the famous Shor algorithm for the factorization
of integers \cite{shor}. Therefore it is important to investigate the effects
of measurements on quantum computers operating nontrivial algorithms.

An interesting example is the quantum algorithm proposed in \cite{georgeot} 
which allows  to simulate the evolution of the kicked rotator on a quantum 
computer. This algorithm essentially uses the Quantum Fourier Transform (QFT)
and controlled phase gates. It realizes one map iteration in a polynomial 
number of quantum gates ($O(n_q^3)$) for a wave vector of size $N=2^{n_q}$.
Here $n_q$ is the number of qubits (two-level quantum systems) onto which 
a kicked rotator wave function is encoded.
For moderate kick amplitudes, this algorithm can be replaced by 
an approximate one which uses all the qubits in an optimal way and performs 
one map iteration in $O(n_q^2)$ elementary gates \cite{andrei}.
In this form the algorithm can simulate complex dynamics, {\it e.g.} the 
Anderson transition, with only a few ($\sim 7$) qubits. 
This makes it accessible for possible future realization on NMR based 
quantum computers. Indeed,  all the elements of the 
algorithm have already been 
implemented on NMR quantum computers \cite{chuang1,cory}.
Therefore it represents an interesting testing ground for the investigation
of the measurement effects on dynamical localization in a quantum computation.

The quantum evolution of the kicked rotator is described by the unitary 
operator $\hat{U}$ acting on the wave function $\psi$ \cite{reviews}: 
\begin{equation}
\overline{\psi} = \hat{U} \psi  =  \hat{U}_k \cdot \hat{U}_T \psi = 
e^{-i k \cos{\hat{\theta}}} 
e^{ -i T \hat{n}^2/2}  \psi .
\label{Uoperator}
\end{equation}
Here $\overline{\psi} $ is the wave function after one map iteration, 
$\hat{U}_k$ represents the effects of the kick in the phase  representation
and $\hat{U}_T$ describes the free rotation in the momentum basis $n$
with $\hat{n} = -i \partial/ \partial\theta$ (we use units with $\hbar=1$).
The dimensionless parameters $k$, $T$ determine  the kick strength and the 
rotation phases, so that the classical limit corresponds to $k \to \infty$, 
$T \to 0$ with the chaos parameter $ K=kT$ constant.
Here we study the regime of dynamical localization 
corresponding to $l \ll N$, where $l \approx k^2/2$ is the localization length
\cite{reviews}.

The quantum algorithm simulating the evolution (\ref{Uoperator}) operates as 
described in \cite{georgeot,andrei}. The wave function $\psi$ 
in the momentum
representation with $N=2^{n_q}$ 
levels is encoded on a quantum computer with $n_q$ qubits. 
In this way $n= -N/2 + j$
where the index $j=0,\ldots,N-1$ is written in the binary 
representation as $j=(a_{1},a_2,\ldots,a_{m},\ldots,a_{n_q})$, with $a_m = 0$
or $1$. As the initial state we choose the momentum eigenstate at $n=n_0=0$,
which can be efficiently prepared from the ground state.
Then, as described in \cite{georgeot} the rotation $\hat{U}_T$ is performed
in $O(n_q^2)$ controlled-phase gates. After that the QFT transforms the 
wave function to $\theta$-representation in $O(n_q^2)$ quantum elementary
 gates (see \cite{chuang}). The kick operator $\hat{U}_k$ is realized in 
$O(n_q^3)$ gates with the help of an additional register \cite{georgeot},
or, for moderate $k$ values, it can be approximately implemented in 
$O(n_q)$ gates without any ancilla, following \cite{andrei}. Finally, 
$\psi$ is transformed back to the momentum basis by the inverse QFT. 
Here  we assume that the gates are implemented without errors,
keeping the analysis of imperfection effects for further studies.

To study the effects of measurements
on the dynamics given by the above algorithm
we assume that after each map iteration (\ref{Uoperator})
a projective measurement of a chosen qubit $m$ is performed. 
The measurement can be represented as the action of two projection operators
$P_0(m)$ and $P_1(m)$ giving for $a_m$
an outcome 0 or 1 with the probability
$||P_0(m) \psi||^2$ or  $||P_1(m) \psi||^2$, respectively.  
The measurement projects the wave function onto one of two subspaces of the 
total Hilbert space, corresponding to momentum states labeled by the indexes 
$j=(a_{1},a_2,\ldots,a_{m},\ldots,a_{n_q})$ with fixed $a_m=0$ or $a_m=1$.
Each subspace is composed of $N/2$ states, given by the direct sum
of $2^{m-1}$ cells of $L=2^{n_q-m}$ consecutive momentum states.
For example, for $m=1$ the most significant qubit is measured and $\psi$ is
projected onto momentum states with $-N/2 \le n < 0$ ($a_1=0$) or 
$ 0 \le n \le N/2-1$ ($a_1=1$); for $m=n_q$ the least significant qubit is 
measured and $\psi$ is projected onto even and odd momentum states.  
Such a measurement is the most natural one for the quantum 
computation process.

Thus, the evolution with measurements is given by the following 
equation for the density matrix $\hat{\rho}$  
\begin{eqnarray}
\hat{\overline{\rho}} = P_0(m)\cdot \hat{U}_k \cdot \hat{U}_T \cdot 
\hat{\rho}  \cdot 
\hat{U}_T^{\dagger} \cdot \hat{U}_k^{\dagger} \cdot P_0(m)+
\nonumber \\
 P_1(m) \cdot \hat{U}_k \cdot \hat{U}_T \cdot
\hat{\rho} \cdot \hat{U}_T^{\dagger} \cdot \hat{U}_k^{\dagger} 
\cdot P_1(m)
\label{densy}
\end{eqnarray}
Here, $\hat{\overline{\rho}}$ is the density matrix after one map iteration
with measurement. 
The direct simulation of this equation is quite costly, since $N^2$ 
components should be iterated. To avoid this difficulty we used the method of
quantum trajectories \cite{qtraj}. In this method for one quantum 
trajectory the  wave function $\psi$ 
evolves  according to (\ref{Uoperator}); after each map iteration
$\psi$ in the momentum representation is projected on the subspaces with
$a_m=0$ or $1$ according to the probability $||P_0(m) \psi||^2$ or 
$||P_1(m) \psi||^2$, respectively.  
After the renormalization, this gives the wave function $\psi_n$ in the 
momentum basis.
The density matrix and 
the expectation values of observables are 
then obtained by averaging over $M$ quantum trajectories.

\begin{figure}[t!]
\includegraphics[width=.99\linewidth]{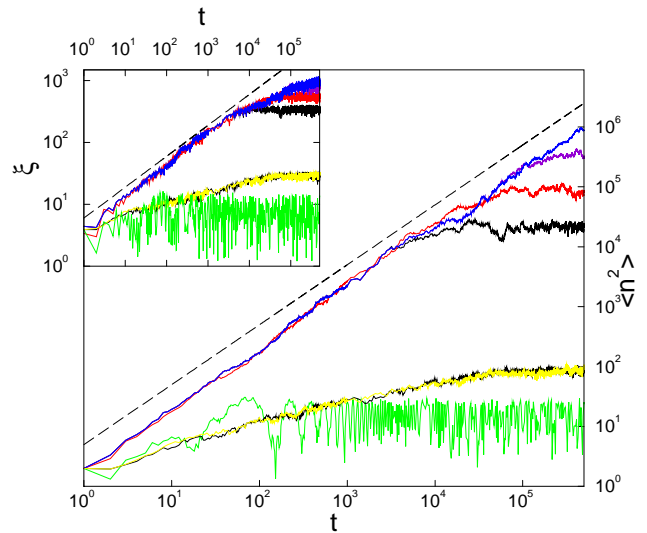}
\vglue -0.4cm
\caption{(Color on line) 
Dependence of the 
second moment $\langle n^2\rangle$ 
on time $t$. Here $T=2$, $k=2$ and color marks the value of $n_q$.
The upper  group of curves corresponds to the measurements of 
the least significant qubit $m=n_q$
for $n_q=$ 12 (blue), 11 (violet), 
10 (red) and 9 (black), from top to bottom. In the lower group 
one of most significant qubits is measured with $m=n_q-8$ for
$n_q=12$ (black) and 9 (yellow)
(data for $n_q=10,11$ give same superimposed
curves and are not shown). The lowest green fluctuating 
 curve is the evolution without measurements.
The dashed line shows the diffusive growth $\langle n^2\rangle \sim t$.
The inset shows the dependence of the IPR $\xi$ on $t$ (colors are 
as in the main plot, the dashed line shows the diffusive growth 
$\xi \sim \sqrt{t}$).  
}
\label{fig1}
\end{figure}

To characterize the quantum evolution with measurements we compute the 
following quantities: the probability distribution $\rho_{nn} \approx 
\langle |\psi_n|^2 \rangle $, obtained
by averaging $|\psi_n|^2$ 
over $M$ quantum trajectories;
the second moment of the probability distribution, 
given by
$\langle \hat{n}^2 \rangle = Tr(\hat{n}^2 \hat{\rho}) \approx
\sum_n n^2 \langle |\psi_n|^2 \rangle$;
the Inverse Participation Ratio (IPR) $\xi = 1/ \sum_n \rho_{nn}^2 
\approx 1/\sum_n |\langle |\psi_n|^2 \rangle|^2$ which determines the number
 of states on which the average probability is distributed.
Within statistical fluctuations these
quantities remain unchanged for a variation of $M$ from 20 to 500
and we represent them for $M=50$.

\begin{figure}[t!]
\includegraphics[width=.8\linewidth]{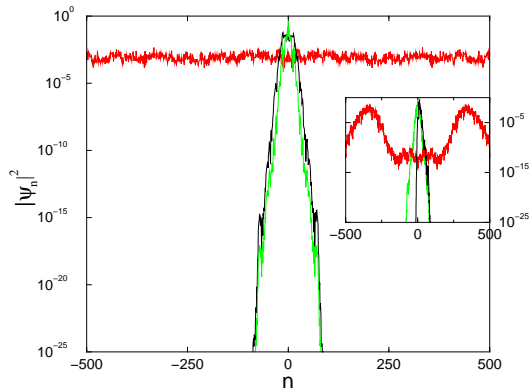}
\vglue -0.4cm
\caption{(Color on line) Probability distribution for $k=2, T=2, n_q=10$
at $t=5 \times 10^5$. Measurements are done for
$m=n_q-8$ that preserves localization (black curve)
and for $m=n_q$ that leads to extended distribution
(red curve). The distribution for evolution without measurements is
shown by green/gray curve. Data are averaged over $50$
quantum trajectories. The inset shows  $|\psi_n|^2$
for a single quantum trajectory (same colors).
}
\label{fig2}
\end{figure}

The dependence of  $\langle \hat{n}^2 \rangle$ and $\xi$ on the number of map
iterations $t$ is displayed in Fig.\ref{fig1}
for different $n_q$ and $m$. 
The probability distribution $\langle |\psi_n|^2 \rangle$ 
for $n_q=10$ is shown in Fig.\ref{fig2}. These data clearly 
show that the measurement of the least significant qubit completely
destroys localization generating a diffusive behavior. 
Indeed, the second moment  $\langle \hat{n}^2 \rangle$ and the IPR $\xi$
grow diffusively (see Fig.\ref{fig1}) up to spread of the probability 
over the whole computational  basis, as shown in Fig.\ref{fig2}. 
The extended distribution $\langle |\psi_n|^2 \rangle$  is formed by a
superposition of probabilities $|\psi_n|^2$ generated by single quantum 
trajectories  (see Fig. \ref{fig2} inset,  which shows that  each 
$|\psi_n|^2$ is relatively narrow). 
On the contrary, the measurement of one of the most significant qubits
does not destroy localization, as clearly illustrated in Figs.\ref{fig1},
\ref{fig2}.  
This striking result 
is very different from the previous studies \cite{graham,kaulakys,facchi} 
where localization was always destroyed by measurements. 

\begin{figure}[t!]
\includegraphics[width=.9\linewidth]{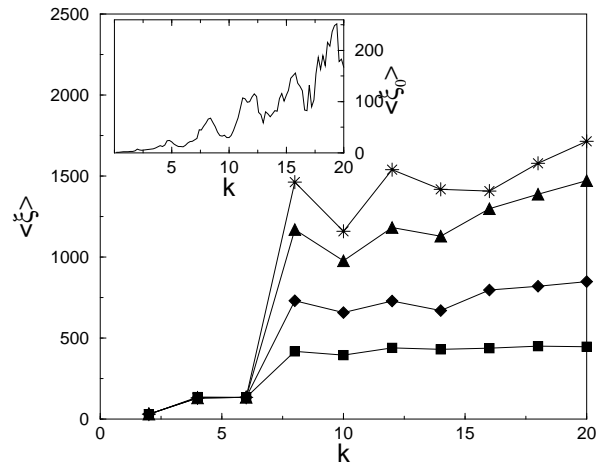}
\vglue -0.4cm
\caption{
Dependence of the averaged IPR $\langle \xi \rangle$ on $k$ for measurements
of one of the most significant qubits $m=n_q-8$, for $n_q=9$ (squares);
$10$ (diamonds); $11$ (triangles); $12$ (stars). IPR values are averaged over 
1000 kicks around $t=5 \times 10^5$; $T=2$.
The inset shows the same dependence in the absence of measurements.
}
\label{fig3}
\end{figure}

To understand the origin of this behavior we investigate the dependence of
the averaged IPR $\langle \xi \rangle$ on the kick amplitude $k$ for 
different number of qubits $n_q$ (see Fig.\ref{fig3}). For $k \le k_c \approx 6
$ the IPR is independent of $n_q$, corresponding to a localized regime. 
On the contrary, for $k > k_c$ the IPR starts to grow with the system size 
$N = 2^{n_q}$, indicating a transition to a delocalized phase. We explain the 
appearance of this transition in the following way. The measurement process
determines the cell size $L=2^{n_q-m}$ inside which the coherence of 
quantum dynamics is preserved. If the unperturbed localization length $l$ 
is much smaller than the cell size, then measurements do not destroy 
dynamical localization. While, if $l \gg L$, the wave function propagates 
over different cells, measurements destroys quantum coherence 
between nearby cells and this leads  to a  diffusive 
propagation over the computational basis. 

According to our data the 
delocalization transition takes place when
\begin{equation}
\xi_0 \approx 2 l \approx k^2 \approx  L/5
\end{equation}
where $\xi_0$ is the IPR for the dynamics without measurements (see the inset
 in Fig.\ref{fig3}). This relation shows that the transition can be obtained
by tuning $k$ at fixed $n_q-m$ or by an appropriate variation of $m$ at 
fixed $k$. Our numerical data confirm this estimate. Indeed, for 
$m=n_q-9$ we obtain that $k=10$ is localized, while 
at $k=12$ delocalization takes
place (data not shown). It is interesting to note that the oscillations 
of $\langle \xi \rangle$ in Fig.\ref{fig3} 
are correlated with the oscillations of $\xi_0$, 
thus confirming that the delocalization border is determined by the unperturbed
localization length $l$ (these oscillations are produced by dynamical 
correlations which affect the classical/quantum diffusion rate related to
the localization length $l$ as discussed in \cite{ds}). 

To study the quantum dynamics at larger time scales we use the random
quantum phase method proposed in \cite{kaulakys}. It is based on the fact that 
after a projection on a given quantum state induced by a measurement the 
quantum phase is not defined. Therefore one can assume that 
states associated to different outcomes of the measurement procedure 
have a random relative quantum phase. Thus, after a measurement of the 
$m$-th qubit the state $|\phi\rangle$ is replaced by 
$e^{i \beta_0} P_0(m)|\phi\rangle + e^{i \beta_1} P_1(m)|\phi\rangle$,
where the phases $\beta_{0,1}$  are random.
This approach allows to reduce  significantly the computational 
cost of the simulation, since it effectively integrates the dynamics 
over many quantum trajectories. 

\begin{figure}[t!]
\includegraphics[width=.9\linewidth]{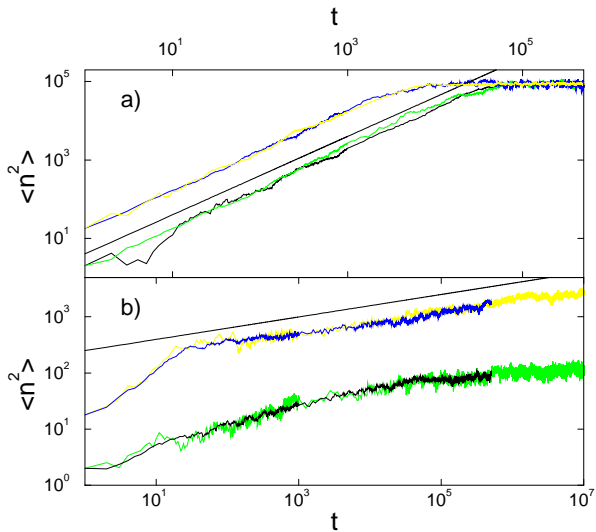}
\vglue -0.4cm
\caption{(Color on line)
Time dependence of the second moment $\langle n^2 \rangle$ 
of the quantum distribution obtained  by  the computation with
quantum trajectories (blue and black curves) and with the random quantum 
phase method (yellow/gray and green/gray curves); $T=2$, $n_q=10$.
Panel a) shows diffusive regime for $k=6$ (upper blue and yellow curves) and
$k=2$ (lower black and green curves), for $m=n_q$; the straight line gives
the diffusive law $\langle n^2 \rangle \sim t$.
Panel b) shows a localized regime for $k=2$ (lower curves) and a near critical
case for $k=6$ (upper curves), for $m=2$; the straight line shows anomalous 
diffusion $\langle n^2 \rangle \sim t^{0.2} $, colors are as in panel a).
}
\label{fig4}
\end{figure}

The comparison of the two computational 
methods is presented in Fig.\ref{fig4}, for diffusive, localized and 
critical regimes. Both methods give consistent results for 
$\langle n^2 \rangle$ (Fig.\ref{fig4}) and the IPR (data not shown).
With the random quantum phase method  we can follow the evolution for  very
large times (up to $t=10^7$) at which localization is still preserved 
(see Fig.\ref{fig4} b). This computational method 
allows also to understand in a better way why localization is not destroyed
by measurements. Indeed, the effects of random phase 
fluctuations $\beta_{0,1}$ appear only at the cell 
boundaries. Hence, for $L \gg l$ they do not affect 
the momentum states located on a distance 
larger than $l$ from edges and localization is preserved
\cite{note}.
We think that the same mechanism qualitatively explains
the results obtained in \cite{sarkar} where it was found that measurements of
a 1/2-spin detector coupled to the kicked rotator do not destroy localization.
In this case the effective cell size $L$ is the total number of rotator 
momentum states and thus localization is preserved since $l \ll L$. 

In conclusion, we studied the effects of measurements on dynamical localization
in a quantum algorithm simulating the kicked rotator.  
Contrary to the common lore the localization is not always destroyed by 
measurements, and a transition from
localized to diffusive dynamics takes place when system parameters are varied.

The authors acknowledge useful discussion with S.Bettelli, 
B.Georgeot. We thank CalMiP in Toulouse 
and IDRIS in Orsay
for access to their supercomputers.
This work was supported in part by the EC projects  RTN QTRANS and
IST-FET EDIQIP
and the NSA and ARDA under ARO contract No. DAAD19-01-1-0553.


\begin{thebibliography}{99}
\bibitem{note} We cannot exclude that this  noise may produce 
hardly observable unlimited diffusion with a rate $\propto \exp(-2L/l)$.
\end{thebibliography}
\end{document}